\documentclass[a4paper]{aa}
\usepackage{graphicx}
\newif\ifAMStwofonts
\newcommand{\be}{\begin{equation}}
\newcommand{\ee}{\end{equation}}
\newcommand{\bea}{\begin{eqnarray}}
\newcommand{\eea}{\end{eqnarray}}

\newcommand{\etal}{ {\rm et al.} }

\begin{document}

\title{Non-radial motion and the NFW profile }


\author{ Morgan Le Delliou
          \inst{1,2}\fnmsep\thanks{delliou@astro,queensu.ca, Morgan.LeDelliou@obs.univ-lyon1.fr}
          \and
          Richard N. Henriksen\inst{1}\fnmsep\thanks{henriksn@astro.queensu.ca}
          }
\authorrunning{M. Le Delliou \and R.N. Henriksen}

   \institute{Queen's University, Kingston, Ontario, K7L 3N6,Canada \\
         \and
            Observatoire de Lyon, Saint Genis-Laval, 69000, France  \\
             }

\date{\today}



\label{firstpage}


   \abstract{
   The self-similar infall model (SSIM) is normally discussed in the
context of radial orbits in spherical symmetry. However it is possible
to retain the spherical symmetry while permitting the particles to
move in Keplerian ellipses, each having the squared angular momentum
peculiar to their `shell'. The
spherical `shell', defined for example by the particles turning at a given radius, then
moves  according to the radial equation of motion of a `shell'
particle. The `shell' itself has no physical existence except as an 
ensemble of particles, but it is convenient to sometimes refer to the shells
since it is they that are followed by a shell code.
In this note we find the distribution of squared angular momentum as a
function of radius that yields the NFW density profile for the final
dark matter halo. It transpires that this distribution is amply 
motivated dimensionally. An effective `lambda' spin
parameter is roughly constant over the shells. We also study the effects
of angular momentum on the relaxation of a dark matter system using 
a three dimensional representation of the relaxed phase space.

     
}

   \maketitle
%

\setlength{\baselineskip}{13pt}
\section{Introduction}
\label{sec:intro}
In the paradigm of Cold Dark Matter (CDM), what have been called universal
density profiles have emerged from collisionless self-gravitating
numerical models in the form 

\begin{equation}
\label{NFW}
\rho \propto r^{-\alpha }\left( 1+\left( \frac{r}{r_{s}}\right) ^{\beta }\right) ^{-\frac{\gamma -\alpha }{\beta }},
\end{equation}

where either \( \alpha =1, \) \( \beta =1, \) \( \gamma =3 \)
(\cite{NFW}) or (\cite{Moore99})
 \( \alpha =1.5, \) \( \beta =1, \) \( \gamma =3 \). Noting the
 importance of mergers in hierarchical clustering Syer \& White
 (\cite{SW98}) argued for this universality as a self-regulation of a
 halo density profile. This is due to tidal stripping and dynamical friction
 acting alternately on merging satellites to flatten a steep profile
 and steepen a flat profile. Independently (\cite{Subraetal99a})  gave
a similar argument wherein a nested sequence
of undigested cores form the profile, each with power law densities
dominating only locally.  However several authors have shown that N-body simulations
reproduce the (\cite{NFW}) profile without the need for clumpy initial
conditions and mergers
(\cite{Aguilar90,Huss99a,Huss99b,Moore99,TittleyCouchman00}). 

The radial Self-Similar Secondary Infall model (SSIM:
\cite{FG84,Moutarde,HW99}) predicts a `one-sided attractor' final 
density profile according to (but see (\cite{HS85}) for a slightly
different dependence on the cosmological spectral index) 

 \begin{equation}
\label{SSIMdenProf}
\frac{\rho }{\rho _{c}}=\left\{ \begin{array}{cc}
\left( \frac{r}{r_{s}}\right) ^{-2}, & n<1\\
\left( \frac{r}{r_{s}}\right) ^{-\frac{3(n+3)}{n+5}}, & n>1
\end{array}\right. .
\end{equation} 
Here $n$ refers to the power spectrum of the primordial cosmological
perturbations piece-wise approximated as $P\propto k^n$. In the range
of acceptable initial conditions the variations in the predicted 
density logarithmic slope are small (\( 2\leq \frac{3(n+3)}{n+5}\leq
\frac{9}{4} \)). Hence the predicted central cusps are too steep
compared to those found in the N-body simulations or indeed compared
to observations (\cite{DeB-etal01}).
In a companion paper (\cite{HLD02}), the SSIM is shown to admit density
inflections at the centre  when finite physical resolution is
accounted for,
but the expected flattening may not be on a large enough scale to reproduce that
of the  N-body simulations. Such resolution effects however may lie
behind the mass resolution dependence in the results of Jing \& Suto (\cite{JingSuto}).

Recently Lokas \& Hoffman (\cite{LH00}) have studied the consequences of
abandoning the self-similar aspect of the SSIM in favour of an
adiabatic invariant calculation, and of introducing a more detailed
form of the initial power spectrum. However the self-similarity has been
found to arise naturally in most shell code simulations (i.e. it is not
an assumption) and in any case the results of Lokas \& Hoffman are very close to those
cited above for the SSIM. The authors conclude by suggesting that it
is prominently the presence of angular momentum that flattens the central cusp.

Indeed numerous authors have emphasized the effect of
 an isotropic velocity dispersion ( thus of non-radial
motion) in the core of collisionless haloes. Thus (\cite{Huss99b})'s
Fig. 17 shows an NFW-type density cusp flattening relative to
the isothermal profile just where the velocity dispersion changes from predominantly
radial to isotropic. This flattening doesn't appear, in (\cite{Huss99a}),
for the case of pure radial force. A similar result was obtained
by (\cite{Tormen97}), and (\cite{Teyssier97}) who also found that singular isothermal
profiles arise during radial infall while isotropic velocity dispersions
 are associated with flatter profiles. Similarly, Moutarde
\etal (\cite{Moutarde}) correlate flat density profiles with higher dimensionality
of the available phase space during infall and they confirm the
 natural development of self-similarity after turn-around.  Thus we
 are motivated to consider a simple extension of the SSIM that
 includes the effects of angular momentum (i.e. of each orbiting
 particle producing in general a smooth velocity `anisotropy' --- the
 limits being purely radial or spherically symmetric in velocity space --- at 
each point on a spherical `shell').

Such a spherical model corresponds strictly neither to reality
 nor to the 3D N-body simulations of cosmological dark halo fields, which display marked non-sphericity (e.g. in \cite{Huss99a,Huss99b,Moutarde,Teyssier97,Tormen97}). Some groups have even proposed a triaxial density profile to replace the spherical (\cite{NFW}) fit (\cite{JingSuto02}). We can regard the
 spherical model as the result of a kind of `coarse-graining' or
 averaging in angle, but it remains unclear in principle whether the averaging
 before the evolution as done here is equivalent to the averaging
 after the evolution as done in the simulations
 (\cite{NFW}). In practice we simply examine to what
 extent the respective profiles agree.

Various  authors (\cite{RydenGunn,WhiteZaritsky,Ryden93,Subraetal99a}, 
\cite{
Subra99b,Sikivie}
\footnote{The focus of (\cite{Sikivie}) was on indentifying velocity streams from non spherically-symmetric
angular momentum distributions. Although the same geometry as used
here was employed,  a (\cite{FG84,Bertschinger85})- type one particle
integration was used which assumes strict  self-similar phase mixing
and so is insensitive to phase
space instability. 
})
have treated  degrees of velocity anisotropy (orbital angular momentum) in the
SSIM and found hints of shallow inner density cusps, and in at least
one case 
links between the NFW inner slope and isotropic velocities
(\cite{RydenGunn})
 were found. However none of these semi-analytic treatments followed
 the system through relaxation by phase mixing and instability
 (\cite{HW99,MH03}) to its final form. In (\cite{Ryden93}) the spherical
 symmetry was broken and the final phase space was studied but the
 particle orbits were confined to poloidal planes. Nevertheless slightly flatter
 spherically averaged density profiles were found which may reflect
 the correlation with higher dimensionality found above (\cite{Moutarde}).  We pursue
 the possible effects of general velocity anisotropy here by the use of a
 {\it spherically-symmetric} shell code, whose only constraint is that
 the angular momentum should be zero averaged over shells.

 In fact the system is made up of particles on elliptical orbits that lie in planes
through the centre of the system and are
isotropically distributed in angle at any point on a sphere. Thus the
vector angular momentum on the sphere is zero. A spherical
`shell' may be defined as the set of particles at a given
radius  that are all at the same phase in their orbits. It might be
the turn-around or apocentric phase for example.
 Subsequently the shell comoves with the same set of particles according to the
radial equation of motion of a particle.    

Because of the  spherical symmetry, angular momentum has to
be introduced to the particles ad hoc, after which it is conserved. 
 This was done in (\cite{WhiteZaritsky})
by introducing a heuristic source term that switches off at turnaround,
or in another context simply by assigning an angular momentum
distribution at turn-around time (\cite{Sikivie}).

In this work, two forms of angular momentum are  assigned at the turn-around
of a given set of particles. It is convenient subsequently to speak of this angular
momentum squared as being assigned to a shell, but the reality is as
described above. In the next section, we will briefly explain our
implementation of angular momentum in the SSIM. The effect
on the density profile will be shown in Sect. \ref{sec:DenNFWSec}.
Sect. \ref{sec:phspace} will focus on other consequences for 
the SSIM's relaxation. A brief discussion regarding mass-angular momentum correlation
will be presented in Sect. \ref{sec:j2MassCorellationSec} before
a concluding discussion (Sect. \ref{sec:summary}).

\renewcommand{\textfraction}{0}
\renewcommand{\topfraction}{1} 
\renewcommand{\bottomfraction}{1}

\section{The SSIM and Particle Angular Momentum}

We seek a natural way of assigning the particle angular momentum on a shell `a priori' so
that our eventual agreement with the NFW density profile should not
simply be an empirical fit. There are two possibilities which seem
rather natural to us. The first one is to assign that distribution of
angular momentum over the `shells' which allows the generalized
self-similarity described by Henriksen (\cite{H89}) to hold until shell
crossing. Unfortunately this self-similarity does not carry through
into the virialized system by way of a constant logarithmic derivative
of the turn-around time with respect to the turn-around radius as in
(\cite{FG84}) or in (\cite{HW99}). In fact rather than yielding a
prediction for the final density profile in terms of the assigned
magnitude of angular momentum as might be anticipated, we find that
 the permitted magnitude is so small that no essential change is produced in the
relaxed density profile. This distribution will be referred
to as the self-similar angular momentum profile or SSAM.

The second distribution that we consider and the one that seems to
produce the most satisfying results physically is simply that given by
dimensional analysis in a spherically symmetric self-gravitating
system. This is simply a power law distribution and will be referred
to as the power law angular momentum profile or PLAM.

\subsection{The SSAM: Self-similarity with angular momentum }  

Following (\cite{H89}) the radial equation of motion may be written in
the `Friedman' form 

\begin{equation}
\label{FriedmanJ2}
\left( d_{\xi }S\right) ^{2}-\frac{1}{S}+\frac{J^{2}}{S^{2}}=-1,
\end{equation}

 where we have defined the 
Lagrangian label (recall that for bound shells, E\( < \)0, hence
the -1 in Eq. (\ref{FriedmanJ2}))

\begin{equation}
\label{TurnAroundRadius}
a=\frac{GM(r)}{\left| E\right| }.
\end{equation}

 The current radius $R(r,t)$ of a particle or spherical shell is the
 scaled form 
\[R=aS(\xi ),\]

 with the self-similar independent variable defined as 
\[\xi =\sqrt{\frac{2GM}{a^{3}}}\left( t-t_{0}(a)\right) ,\]

and \( t_{0}(a) \) is  the turn-around time, chosen so that at turn-around
\( \xi =0 \). 
Maintaining the self-similarity of Eq.(\ref{FriedmanJ2})
requires $J^2$ to be constant, which yields a constraint on the square
of the `shell angular momentum' $j^2$ through the definition 
\begin{equation}
\label{SSangMconst}
J^{2}=\frac{j^{2}}{2GMa}.
\end{equation}

This quantity may also be written using equation
 (\ref{TurnAroundRadius}) in the form 
\begin{equation}
J=\frac{j|E|^{1/2}}{\sqrt{2}GM},
\end{equation}
which shows it to be, but for a factor $\sqrt{2}$, the `local' spin parameter 
$\lambda$ (\cite{Peebles93}) expressed in terms of specific angular
momentum and energy. The mass however is the entire mass inside a
given shell rather than simply that of the particles constituting the
shell. 

The solution to Eq. (\ref{FriedmanJ2})  before shell crossing and
up to turn-around is 
\begin{equation}
\xi+\pi/4=-\sqrt{S-(S^2+J^2)}+\frac{1}{2}\arcsin{\frac{2S-1}{\sqrt{1-4J^2}}},
\label{scalefactor}
\end{equation}
which gives $\xi=0$ at the turn-around scale $S_T$ since there 
\be
S-(S^2+J^2)=0,
\ee
and thus 
\be
2S_T-1=\sqrt{1-4J^2}.
\ee

At $t=0$ Eq. (\ref{scalefactor}) yields the relation 
\be
\sqrt{\frac{2GM}{a^3}}t_o(a)=\frac{\pi}{4}+\sqrt{S_o-(S_o^2+J^2)}+\frac{1}{2}\arcsin{\frac{1-2S_o}{\sqrt{1-4J^2}}},
\label{eq:xi0turnaround} \ee
where $S_o\equiv r/a$, the ratio of the initial radius of a `shell'
and its Lagrangian label. Once this latter ratio is determined the
turn-around time $t_o$ may be found as a function of $a$. This ratio
follows from Eqs. (\ref{TurnAroundRadius}) and
(\ref{SSangMconst}) once $E(r)$ and $M(r)$ are given initially, as
does also the required $j^2(r)$. Thus this requirement of
self-similarity {\it before} shell crossing leads to a natural
initial correlation between $j^2$ and the mass inside a shell.

 The expression for the initial mass inside radius $r$ for a constant density background
perturbed by a power law $\rho=\rho_b(1+x^{-\epsilon})$ is 
\begin{equation}
\label{SSIMinitMass}
M(r)=M_{f}x^{3}\left( 1+\frac{x^{-\epsilon }}{q}\right) 
\end{equation}
 where q is defined in terms of the power-law index $\epsilon$ of the initial density perturbations
\[q=(1-\frac{\epsilon }{3})\]
, and we adopt\[\rho =\rho _{b}\left( 1+\lambda .r^{-\epsilon }\right) ,\]
 \( \lambda  \) being constant. The scaled radius $x\equiv r/r_f$ is
 defined in terms of a
fiducial radius (\( r_{f}=\left( \lambda \right) ^{\frac{1}{\epsilon }} \))
and the  fiducial mass is introduced as \[M_{f}=\frac{4\pi }{3}\rho
 _{b}r_{f}^{3}.\] The background density is \( \rho _{b} \). 

We calculate the initial specific energy of a shell in a de Sitter
cosmology after the addition of the angular momentum as 
\begin{equation}
\label{SSIMinitE}
E(r)=\frac{H^{2}r^{2}}{2}-\frac{GM(r)}{r}+\frac{j^{2}}{2r^{2}}=-\frac{4\pi G\rho _{b}\lambda }{(3-\epsilon )}r^{(2-\epsilon )}+\frac{j^{2}}{2r^{2}}.
\end{equation}

Now combining Eqs. (\ref{SSIMinitMass}) and (\ref{SSIMinitE}) in the
definition (\ref{TurnAroundRadius}), one finds for example
\be
\frac{x}{a}=\frac{1+\sqrt{1-4J^2(1+q
    x^\epsilon)^2}}{2(1+qx^\epsilon)},
\label{eq:x/a}
\ee

 which yields using Eq.(\ref{SSangMconst}) the expression for the
self-similar compatible angular momentum distribution in the SSIM
\footnote{The approach from the force equation in (\cite{Sikivie}) yields a slightly
different self-similar constraint for the angular momentum:\[
j^{2}_{Sikivie}=\frac{J^{2}_{Sikivie}S^{4}_{T}a^{4}}{t_{o}(a)^{2}}=\frac{2J^{2}_{Sikivie}S^{4}_{T}GMa}{\xi _{a}^{2}}=\frac{j^{2}}{\xi _{a}^{2}},\]
where \( \xi _{a}(x)=\left. \xi \right| _{t=0} \) is obtained with
the combination of Eqs. (\ref{eq:xi0turnaround}), (\ref{eq:x/a})
and \( S=x/a \) and we identify our conventions with \( J^{2}_{Sikivie}S_{T}^{4}\equiv J^{2}. \)
}
 as

\be
\frac{j^2}{2GM_fr_f}=\frac{2J^2}{q}~x^{(4-\epsilon)}~\frac{(1+qx^\epsilon)^2}{1+\sqrt{1-4J^2(1+qx^\epsilon)^2}}.
\label{eq:angssam}
\ee

Now it is immediately clear from these expressions that the imposed
self-similarity can not hold beyond $x_{max}=\left((1-2J)/(2Jq)\right)^{1/\epsilon}$. The
specific energy at this radius follows from Eqs.
(\ref{TurnAroundRadius}), (\ref{eq:x/a}), and (\ref{SSIMinitMass}) as 
\be
E=-\frac{GM_f}{2qr_f}x_{max}^{2-\epsilon},
\ee
and is not zero. Thus the self-similarity itself imposes an outer
cut-off on the system. This does not in itself present a problem
peculiar to the imposed self-similarity, since of course for the shallow initial density
profile ($\epsilon<2$) the whole Universe is bound to the halo unless
an arbitrary cut-off is imposed. Moreover even the steep density
profile requires an arbitrary outer cut-off to define a finite mass system.  

The mass at this self-similar cut-off is 
fortunately large compared to the fiducial mass as it becomes (\ref{SSIMinitMass})
\be
M_{max}=\frac{M_f}{(2Jq)^{(3/\epsilon)}}(1-2J)^{(3/\epsilon-1)},
\ee
and  $J$ is normally in the range $10^{-3}$ to $10^{-4}$ with
$\epsilon\approx 2$ in our simulations.   Larger
values (one might envisage adding a spin parameter value up to
$\approx 3.5\times 10^{-2}$)  did not permit a sufficiently
large  extent to allow the self-similar virialized phase to be
well resolved in time (see $x_{max}$ above). The cases that
avoided this numerical limit {\it did not show any significant
  deviation in density or phase space from the halos formed by purely
  radial infall.} We therefore do not dwell further on this
distribution when discussing the density profiles.

\subsection{PLAM: The Power Law Distribution of Angular Momentum}

In this section we consider the simplest initial distribution of angular
momentum (that is a correlation with the initial mass as above) that may be expected on dimensional grounds. This is the
Keplerian form (the specific rotational kinetic energy is a fixed
fraction $J^2$ of the specific binding energy)
\be
\frac{\textrm{j}^{2}}{2r^{2}}=\frac{GM(r)J^{2}}{r}
\label{eq:jkep}
\ee
where $J^2$ is again constant. As a consequence all of the equations
of the previous section (including the solution for the scale factor) may be applied simply by replacing $J^2$ in
those equations by $J^2x/a$ and noting that in this case the previous
procedure leads to 
\be
\frac{x}{a}=-J^2+\frac{1}{1+qx^\epsilon}.
\label{eq:kepx/a}
\ee
The motion before shell crossing is now no longer `self-similar' in
the sense of the previous section, but in fact we can see that for
$J^2x/a$ small the situation at turn-around is little different from
the purely radial motion. It is during the subsequent re-collapse that
angular momentum is significant. One of our
objectives is to see what form this initial correlation takes in the final
relaxed halo so that it may be compared with the $\Lambda$ CDM results
of Bullock \etal (\cite{Bullock01}). 

Once again the angular momentum distribution can not be maintained
beyond an outer scale $x_{max}$ where now 
\be
x_{max}=\left(\frac{1-J^2}{qJ^2}\right)^{1/\epsilon},
\ee
although with $J^2$ small this yields a larger dynamic range than before.
Eqs. (\ref{eq:jkep}) and 
(\ref{SSIMinitMass}) now yield the explicit
initial angular momentum distribution as 
\be
\frac{j^2}{2GM_f r_f}=\frac{J^2}{q}~x^{(4-\epsilon)}(1+qx^\epsilon).
\label{eq:explj2}  
\ee
We observe that at large $qx^{\epsilon}$ this distribution tends to
one for which there is a constant angular velocity as a function of initial
radius. Moreover while the coefficient $J^2(1+qx^\epsilon)$ is
constrained to be always smaller than one, $x_{max}^{(4-\epsilon)}$ may
be large. It is for this reason that $j^2$ may be large enough in this
case to substantially alter the nature of the final collapsed halo.

In the numerical work below we present examples for $\epsilon=1.5$,
the `shallow' case and for $\epsilon=2.5$, the `steep' case. For the
SSAM $J$ is essentially the specific spin parameter ($\times 1/\sqrt{2}$) and calculations
were carried out with $J\approx 1.7\times 10^{-3}$ for the shallow
case and with $J\approx 6.8\times 10^{-5}$ for the steep case. These
values are rather small compared to the median value of $\lambda =.05$
reported in (\cite{Peebles93}). This shows clearly the limitations
imposed by this angular momentum distribution together with the
constraint of forming a core.

For the PLAM distribution, $J^2$ is actually rather larger than the spin
parameter wherever $x/a$ is small. The values used for $J^2$ were
$ 10^{-3}$ for the shallow case and $\approx
9\times 10^{-3}$ in the steep case. These are much closer to the
values for the spin parameter found in cosmological simulations. If
$J^2$ is taken much larger than these values  the outermost shells
did not reach the center. In this latter event the core displayed
isolated phase mixing `islands' (\cite{Sikivie}).



\section {Numerical Evolution of the PLAM
  Distribution to the NFW Profile}\label{sec:DenNFWSec}

\subsection{Establishment of the NFW profile}

\begin{figure*}
{\centering \includegraphics[width=\textwidth]{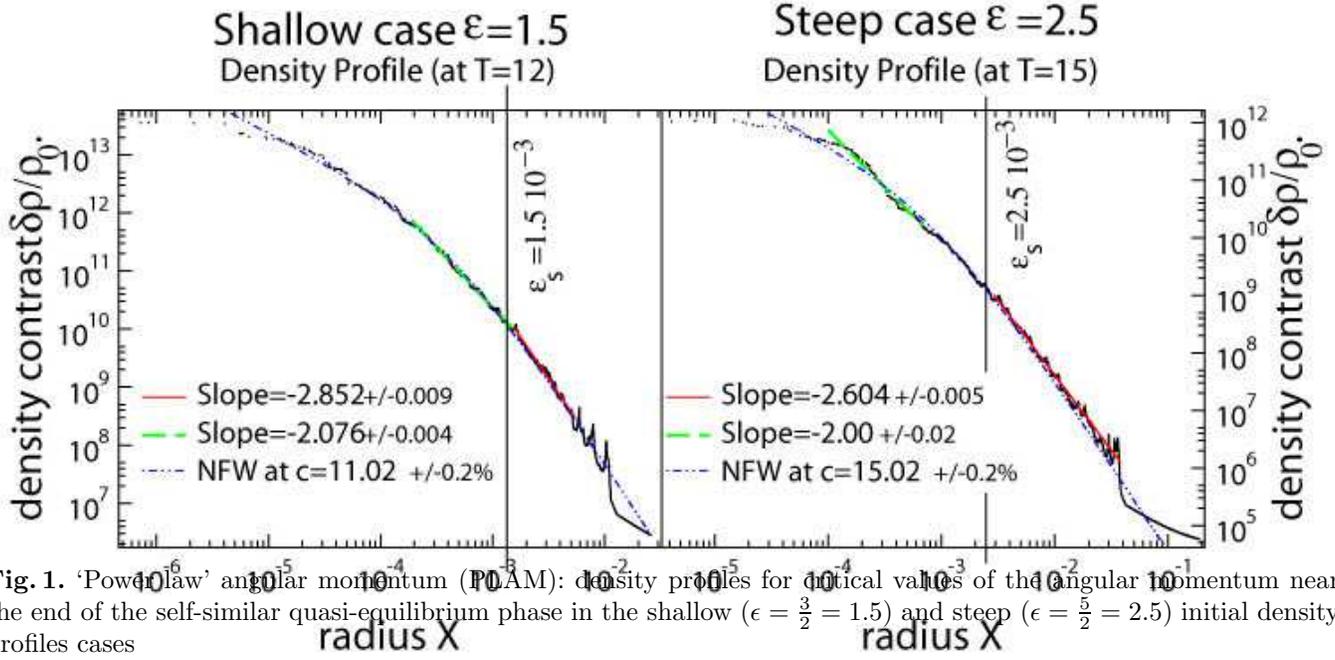} \par}
\vspace{-1.8cm}

\caption{\label{fig:pwrDen}`Power law' angular momentum (PLAM): density profiles
for critical values of the angular momentum near the end of the self-similar
quasi-equilibrium phase in the shallow (\protect\( \epsilon =\frac{3}{2}=1.5\protect \))
and steep (\protect\( \epsilon =\frac{5}{2}=2.5\protect \)) initial
density profiles cases}
\end{figure*}

We use a shell code with a semi-analytic treatment of the energy
calculation (\cite{LD01}) to follow the development of the dark matter
halo through the shell crossing phase. This code has been well tested
elsewhere (\cite{LD01}) and is known to reproduce standard results (\cite{HW99}). 

As was indicated above the SSAM distribution did not produce density
profiles that were significantly different from those found in pure
radial infall. In particular the relations between the power law index
of the initial density perturbation $\epsilon$ and the final density
power law in the bound core were the same as those found for radial
infall (\cite{FG84, HW99}). Thus we do not present these results
here.

The PLAM initial distribution with the parameters given above evolved
to cores having the density profiles shown in (Fig. \ref{fig:pwrDen}).
The time indicated on the figures is that defined in (\cite{HW99}) and refers to 
the logarithmic time near the end of the self-similar infall `equilibrium' 
phase. 

The figure shows that both in the steep and shallow case the profile
is no longer well fit by a power law even in the intermediate
regions. Rather there is a smoothly varying convexity in the
logarithmic slope that is well fit
by the indicated NFW profile {\it except in the most central regions}.
There a flattening occurs that is more pronounced than is accounted
for by the NFW profile.  Flattening is anticipated relative to the self-similar slope
on general thermodynamic grounds in the coarse-graining study of
(\cite{HLD02}), and the presence of a central point mass (perhaps
imitated here by numerically smoothed density cusps) can produce
such flattened cusps (\cite{NakMak99}). Unfortunately our numerical resolution is too poor in
this core region to say that this flattening is a physically significant
result.

We also show piecewise power law fits in
various regions of the figure in order to emphasize the inadequacy of a global power law
fit. One sees moreover that the apparent slope does tend toward
$r^{-3}$ in the external regions. The theoretical power law slope for
undisturbed radial infall is $2.0$ for the shallow case and $\approx
2.14$ for the steep case. Neither of these values fits 
the whole density profile of these cores.
 
We note that the force `smoothing' length $\epsilon_s$ occurs near the
mid-point of the indicated spatial range. This does not affect the
validity of the density profile inside this radius however as was
established numerically and by coarse graining in (\cite{HLD02})
for pure radial infall. Their interpretation of the results in terms
of `turn-round' relaxation of the shells is unchanged by the present extention of phase
space dimensionality since, as seen in Sect. \ref{sec:phspace},
relaxation still takes place mostly outside of the `smoothing' length
(\cite{LD01}). However even if the relaxation argument still holds,
it cannot be made to predict the exact slope of the NFW profile. We shall
examine in detail the NFW profile dependence on resolution in Sect.
\ref{sec:ResNFWSec}.

Our results ought to be compared to the results of the similar
study of Hozumi \etal (\cite{HBF2000}). These authors also investigate
the effect of smoothly 
anisotropic velocity distributions (but  with zero net
angular momentum on shells, just as in our case) on the density profile. Instead
of using a shell code however, they integrate the CBE directly
starting from non-equilibrium power law density
distributions. Moreover they initiate their calculation with an `ad hoc'  distribution
function that is simply a Gaussian in each of the radial and azimuthal
velocities, each Gaussian having different dispersions. 
In contrast we have started our simulations in this work from actual cosmological
initial conditions radially, on which we have imposed rather natural
distributions of angular momentum. The effect is that our collapse
begins from a well-defined surface in phase space.

Despite the differences in the initial conditions, we observe that the
conclusions drawn in the two works are rather similar. Their parameter
$2\alpha$ is essentially our parameter $J^{-2}$ (our radial specific
kinetic energy is essentially $GM/r$) which we have taken in
the range 
$\ 100$- $1000$ for the favoured PLAM. Although the largest value used
in their paper is $2\alpha=20$, because our ratio of kinetic energy to
binding energy ($\approx (1+J^2)/(1+\delta M/M)$) is much closer to
unity than is theirs, the calculations are more comparable than it
might seem. Hozumi \etal have not fitted their profiles with a NFW
curve, but their
conclusions about the central flattening resemble ours. This suggests
that particle angular momentum  is indeed playing a role in determining the
density profiles of dark matter halos, and hence of galaxies.

We conclude then in this section that the PLAM of Eq. (\ref{eq:explj2}) does yield the
NFW density profile over most of a dark matter halo. Moreover the required
amplitude of the angular momentum does not need to exceed that likely
to be provided by tidal interactions. The PLAM places the embryonic
halo particles 
on a surface in phase space which may therefore reveal something of
the tidal fields they experience near turn-around.

\subsection{NFW and resolution}

\label{sec:ResNFWSec}

\begin{figure*}
{\par\centering \resizebox*{!}{0.4\textheight}{\rotatebox{-90}{\includegraphics{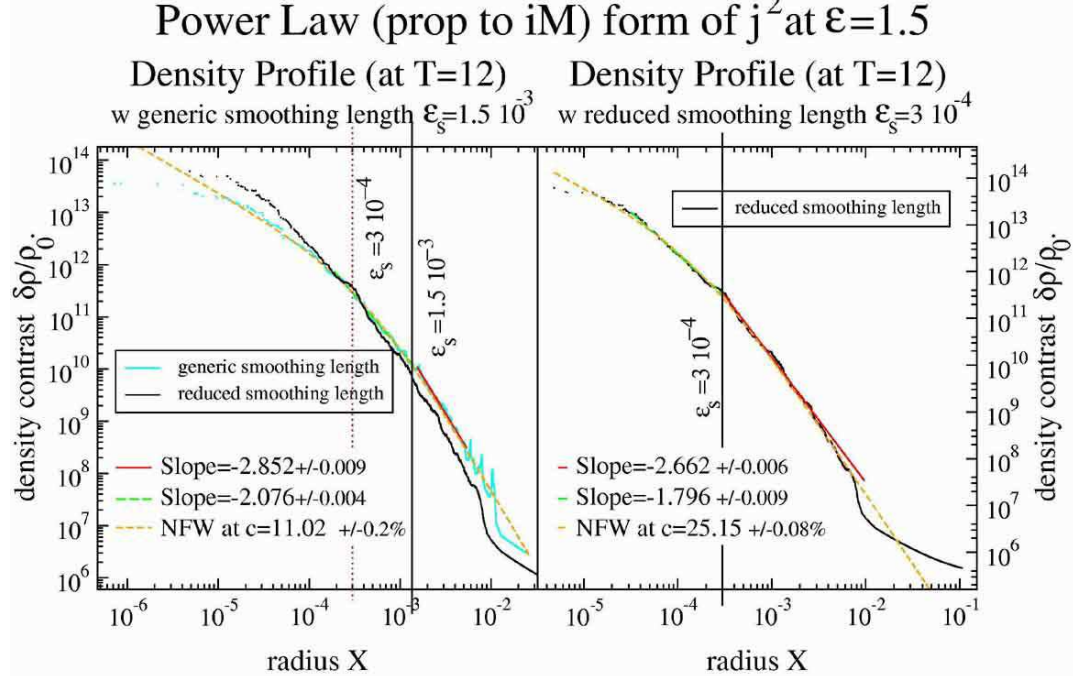}}} \par}

\caption{\label{fig:smlEpsDen1.5}Smaller smoothing length \protect\( \epsilon _{s}\protect \)
test run: density profiles for critical values of the angular momentum near
the end of the self-similar quasi-equilibrium phase in the shallow (\protect\( \epsilon =\frac{3}{2}=1.5\protect \))
initial density profiles case using the `power law' form of angular momentum}
\end{figure*}

The motivation in (\cite{Moore99}) for proposing an alternate universal
profile to the NFW profile is based on their observation of a resolution
dependence of the inner slope. This resolution problem was also pointed
out in the studies by (\cite{JingSuto}). (\cite{Moore99}) claim that
their results show that the inner slope converges towards their proposed profile. 

We have tested our results under changes in resolution using two different `smoothing' lengths.

Similarly to the results of (\cite{Moore99}), an increase in resolution
in our model steepens the profile. This can be measured by the increase
of the concentration factor c%
\footnote{The NFW profile in terms of critical density, density and radial scales
reads\[
\frac{\rho }{\rho _{c}}=\frac{\delta _{c}}{\frac{r}{r_{s}}\left( 1+\frac{r}{r_{s}}\right) ^{2}},\]
with \( r_{s}=r_{200}/c \) defining the concentration factor and
the two scales correlated through (see (\cite{NFW}))\[
\delta _{c}=\frac{c^{3}}{\left[ \ln (1+c)-\frac{c}{1+c}\right] }.\]

} when the `smoothing' length is reduced, as seen on Fig. \ref{fig:smlEpsDen1.5}.
Nevertheless, contrary to (\cite{Moore99}), the NFW profile is still
 the best fit, provided the appropriate change in concentration factor
has been performed.  

We interpret these results in terms of the centrifugal
acceleration at a given radius between the two `smoothing' lengths: thus
its regularised expression in the model reads\[
a_{c}=\frac{j^{2}r}{(r^{2}+\epsilon _{s}^{2})^{2}}.\]
Taking its value at radius \( r_{0}\ll 1 \) with \( r_{0}\in \left[ \epsilon _{s_{2}};\epsilon _{s_{1}}\right]  \),
and considering the dominant terms, one can see that\[
\left. \begin{array}{cc}
for\; \epsilon _{s_{1}} & a_{c_{1}}\propto j^{2}r_{0}\\
for\; \epsilon _{s_{2}} & a_{c_{2}}\propto \frac{j^{2}}{r_{0}^{3}}
\end{array}\right\} \Rightarrow a_{c_{1}}=a_{c_{2}}r_{0}^{4},\]
which yields \( a_{c_{2}}/a_{c_{1}}=1/r_{0}^{4}\gg 1\Rightarrow a_{c_{2}}\gg a_{c_{1}}. \)
In other words, to diminish \( \epsilon _{s} \) depletes particles
from the centre.

Therefore resolution appears to fix the magnitude of the
NFW profile's concentration parameter c. Nevertheless, we have established
that the addition of angular momentum to CDM haloes 
transforms
the single power-law density profile into two slope NFW type profile.

\section{Phase space structure of the dark matter
  halo}\label{sec:phspace}

\subsection{Virial ratio and phase space projection}

\begin{figure}
{\centering \resizebox*{1\columnwidth}{!}{\includegraphics{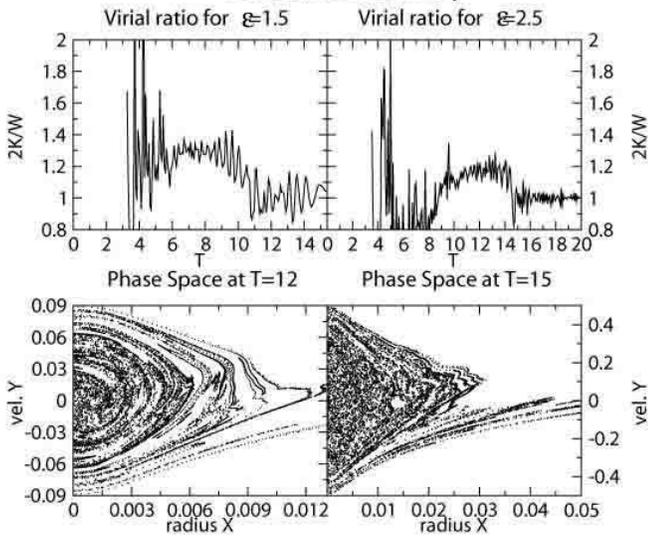}} \par}

\caption{\label{fig:ssj2VirialPhSp}SSAM: virial ratios and phase space projections
in the radius/radial velocity plane near the end of the self-similar
quasi-equilibrium phase for critical values of the angular momentum
in the shallow (\protect\( \epsilon =\frac{3}{2}=1.5\protect \))
and steep (\protect\( \epsilon =\frac{5}{2}=2.5\protect \)) initial
density profiles cases}
\end{figure}

\begin{figure}
{\centering \resizebox*{1\columnwidth}{!}{\includegraphics{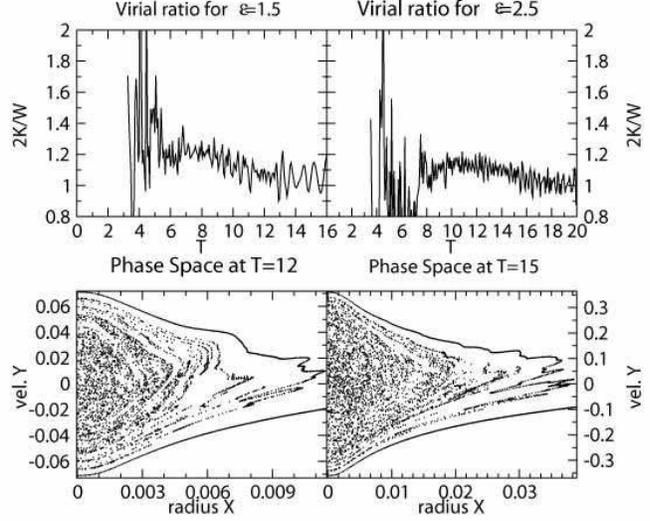}} \par}

\caption{\label{fig:pwrViralPhSp}`Power law' angular momentum (PLAM): virial
ratios and phase space projections in the radius/radial velocity plane
near the end of the self-similar quasi-equilibrium phase for critical
values of the angular momentum in the shallow (\protect\( \epsilon =\frac{3}{2}=1.5\protect \))
and steep (\protect\( \epsilon =\frac{5}{2}=2.5\protect \)) initial
density profiles cases}
\end{figure}

We begin by considering the $r$-$v_r$ projection of phase space and the
corresponding virial ratios, for comparison with the well-known radial
results (\cite{HW99}). These are shown in Figs.
\ref{fig:ssj2VirialPhSp} and \ref{fig:pwrViralPhSp} for the SSAM and
the PLAM respectively. In each figure the `shallow' and `steep'
initial profiles are displayed, although these in fact show little
difference in behaviour. In our simulations we used  `equal mass
modeling' of the `shells'. 
  
In (\cite{HW99}), the shell modeling used smaller masses for inner 
shells than for outer shells in order to improve the dynamical mass resolution
at the beginning of the infall. This allowed for the SSIM's quasi-equilibrium
self similar state to be achieved almost as soon as the core forms,
to the detriment of an accurate phase space description that is our
objective here. The use of constant
mass shells leads to a relatively coarser mass resolution in the central part
of the halo, which part corresponds to the set of shells that first
forms the core,  and so the quasi-equilibrium state is not
achieved as gracefully. In fact this choice results in a poor modeling
of the  constant self-similar mass flux needed to maintain
self-similarity during the accretion of innermost shells into the core.
Each new shell acts almost as an overdensity perturbation to the previously
established core and disturbs it from self-similar equilibrium until
it is `digested' (\cite{LD01}).  Thus the equal mass modeling simulation requires an initial period
of stabilization to reach the self-similar quasi-equilibrium. This
period should be roughly the time to accrete  the shell of the unequal mass modeling
simulation that possesses the same mass as the constant mass used
here. This conjecture was successfully tested.

A major physical difference with the radial SSIM appears during the  transition
from the self-similar equilibrium to a virialized isolated system, for which the virial
ratio is unity. 
 There is a visible smoothing compared
with the radial SSIM's sharp transition: instead of falling almost instantly
from the self-similar value to  unity when the 
last shell falls in, the presence of angular momentum produces
instead a slow decrease from the self-similar value to that of the isolated
system. This begins earlier than in the radial SSIM case (as measured
by $T$) and finishes later.

The phase space projections (the lower panels of figures \ref{fig:ssj2VirialPhSp} and
\ref{fig:pwrViralPhSp} display the phase space distribution
of shells in the radius-radial velocity plane) reveal that there remains a stream of
outer particles for which the angular momentum is so large that they
will never fall into the core (their rotational kinetic energy makes
them unbound). Indeed, since the initial angular momentum distribution for
the particles is
monotonically increasing with radius, and since the angular momentum is conserved
exactly throughout the simulations including the initial radial ordering
 until shells reach the core; so near the end of the self-similar
phase, the shells with increasingly large angular momentum are contributing
to the mass flux. Given higher and higher angular
momentum, there is a point at which angular momentum induces an inner
turn around radius at the size of the self-similar core. Particles with
 smaller angular momentum will be able to enter the core but with
a reduced radial velocity compared with the purely radial radial
SSIM. This gradual
extinction of the mass flux due to increasing particle angular
momentum (velocity anisotropy) gradually
shifts the system from its intermediate quasi-static phase to its
final virialized 
phase.

Comparing the phase space projections of the SSAM and the PLAM in
Figs. \ref{fig:ssj2VirialPhSp} and \ref{fig:pwrViralPhSp} respectively
we note that the SSAM seems less homogeneous. The exploration of the
3D phase space in Sect. \ref{sec:3DPhSp} will confirm the impression
that the SSAM is less relaxed in phase space during the accretion
phase.

\subsection{resolution effects}

Considering the way resolution affects the virial ratio and \( r \)-\( v_{r} \)
projection of phase space, the previous remarks on the system's relaxation
can be extended. 

\begin{figure}
{\par\centering \resizebox*{1\columnwidth}{!}{\includegraphics{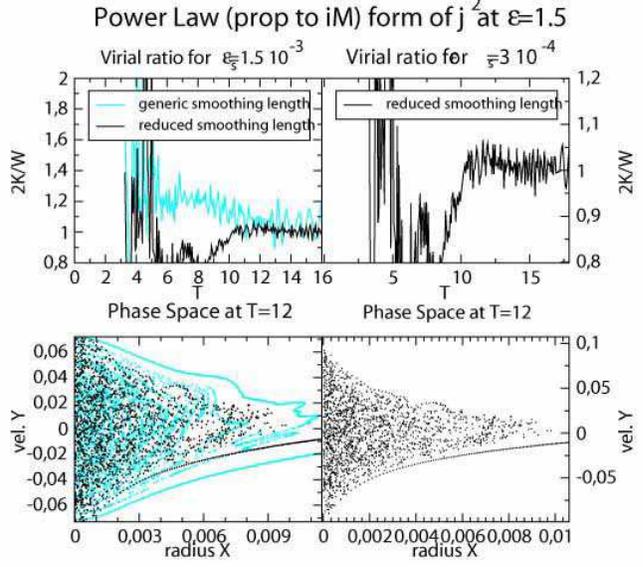}} \par}

\caption{\label{fig:smlEpsVirialPhSp1.5}Smaller smoothing length \protect\( \epsilon _{s}\protect \)
test run: virial ratios and phase space projections in the radius/radial velocity
plane near the end of the self-similar quasi-equilibrium phase for critical
values of the angular momentum in the shallow (\protect\( \epsilon =\frac{3}{2}=1.5\protect \))
initial density profiles cases using the `power law' form of angular momentum}
\end{figure}

The previous stabilisation delay for equal mass `shell' modeling is
increased when using the reduced `smoothing' length. Here this period
encompasses almost all of the self-similar phase (Fig. \ref{fig:smlEpsVirialPhSp1.5}'s
upper panels). Nevertheless, the right panel indicates that there
still is an interval where the virial ratio is slowly decreasing from
a higher value than 1.

We interpret this interval as due to the increase in the maximum
acceleration felt by shells in the central parts of the halo: the scattering of
 shells
is then much stronger, which makes it more difficult for the system to
settle down. To minimise this inertial noise, the mass resolution should
follow the `smoothing' length reduction. We adopted an optimum balance
between the mass, force and time resolutions found by trial and
  error. 

It is remarkable that the noise in the virial ratio (Fig. \ref{fig:smlEpsVirialPhSp1.5})
is diminished with the `smoothing' length: if on one hand the equilibrium
is achieved slower, it is on the other hand of a more stable final nature.
Thus even if the mass flux resolution is not sufficient to account for
the stabilisation delay, it still provides a good basis of understanding
the relaxed system. 

Adding to this picture, the phase space of the smaller \( \epsilon _{s} \)
simulation appears more relaxed than its larger \( \epsilon _{s} \)
counterpart; the relaxation region at the edges of the system restricts
itself to the first outer stream and all traces of the phase space
mixing sheets washes out (Fig. \ref{fig:smlEpsVirialPhSp1.5}'s
lower right and left panels). Such relaxation explains why the smaller
\( \epsilon _{s} \) simulation's phase space is slightly wider in
radial velocity and narrower in radius. This also shows in the more
concentrated smaller \( \epsilon _{s} \) NFW fit (Fig. \ref{fig:smlEpsDen1.5}).

In the light of N-body studies such as that of Knebe \etal (\cite{Knebeetal00}),
inaccuracy in the dynamics from two body scattering excludes `smoothing'
lengths much smaller than that used in our regular simulations, at given
mass resolution. The mass resolution limitations on accuracy as well
as the smoothing-length-time-step relation shed light on the low smoothing
length cut off%
\footnote{An example calculation: the constant mass of one shell is given in
the simulation's unit as \( m_{1shell}=2.10^{-2}. \) The maximum
density contrast on Fig. \ref{fig:smlEpsDen1.5} can be taken as
\( \frac{\delta \rho }{\rho _{0}}=4.10^{13}. \) Because of its high
value, the density contrast, denoted \( \delta  \) in this note, can
be identified with the density itself\[
\frac{\delta \rho }{\rho _{0}}=4.10^{13}\gg 1\Leftrightarrow \frac{\delta \rho }{\rho _{0}}\equiv \delta \simeq \rho .\]
Thus, the volume of innermost shells can be evaluated as\[
\delta \simeq \rho =\frac{m_{1shell}}{V_{1shell}}\Rightarrow V_{1shell}\simeq \frac{m_{1shell}}{\delta },\]
so the characteristic length scale of a shell in the centre is given
by \[
L_{1shell}\simeq ^{3}\sqrt{\frac{m_{1shell}}{\delta }}=^{3}\sqrt{\frac{2.10^{-2}}{4.10^{13}}}=^{3}\sqrt{5.10^{-16}}=^{3}\sqrt{0.5}10^{-5}\simeq 7.9\times 10^{-6}.\]
 
}. 

The nature of relaxation in the SSIM leads to the density profile
being trustworthy even below its `smoothing' length, but only a few
times above its mass resolution characteristic length. This is caused
by the fact that relaxation in the SSIM essentially occurs in the
few dynamical times that particles are freshly incorporated into the
self-similar core, mainly around the secondary turnaround radii, i.e.
near the radial boundary of the core.

All of the phase space maps presented are measured at T=12 which corresponds
to a clear end of the self-similar infall phase and beginning of the
isolated system virialised phase.

\subsection{Three Dimensional Phase Space}
\label{sec:3DPhSp}

This section explores the nature of the mass distribution in the radius-radial
velocity-angular momentum phase space for systems with a shallow (\( \epsilon =\frac{3}{2} \))
initial density profile. The SSAM and PLAM models are both presented here
for comparison purposes.
All of these phase space maps are measured at T=12 which corresponds
to a clear end of the self-similar infall phase and the beginning of the
isolated system virialized phase.  To get a complete
sense of the topology of the winding and relaxation of the original
Liouville stream of Hubble flow shells, we use tilted projections of the
(X,Y,\( j^{2} \)) space. We recall that initially, with the power law
distribution of angular momentum and the Hubble radial velocity, the
particles lie on a curve in this phase space. 

\begin{figure*}
{\centering \resizebox*{0.495\textwidth}{!}{\rotatebox{-90}{\includegraphics{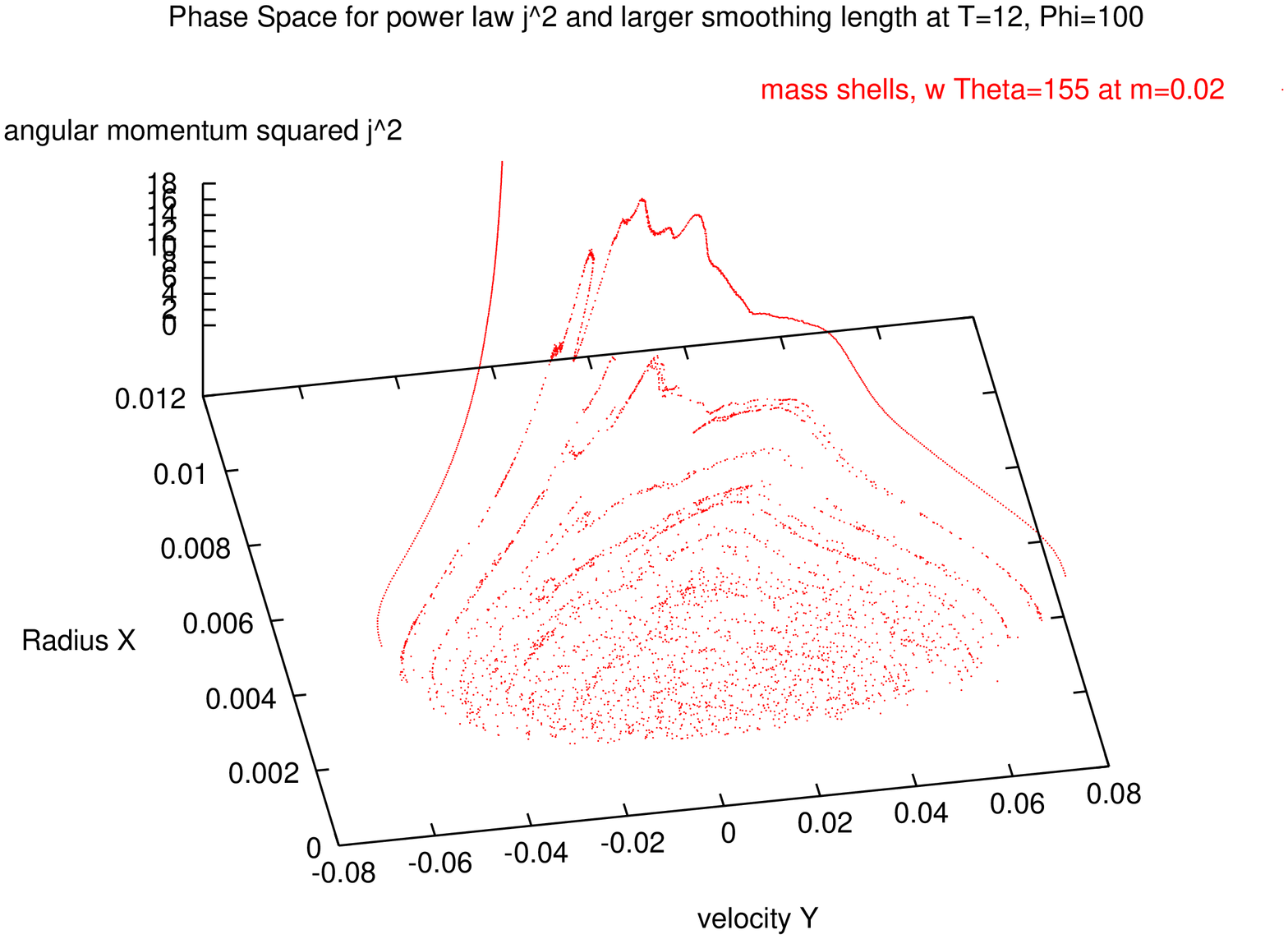}}} \resizebox*{0.495\textwidth}{!}{\rotatebox{-90}{\includegraphics{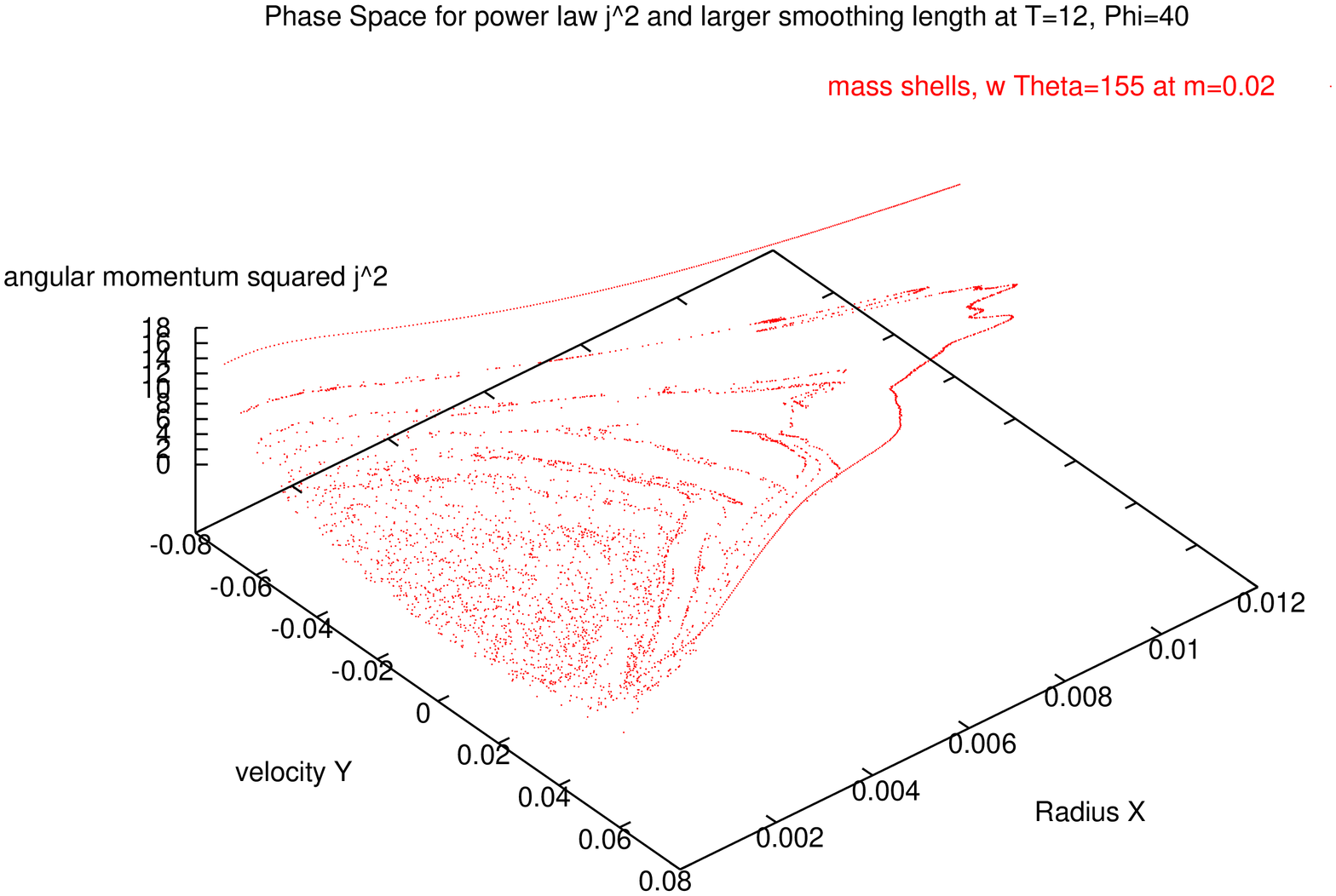}}} \par}

{\centering \resizebox*{0.496\textwidth}{!}{\rotatebox{-90}{\includegraphics{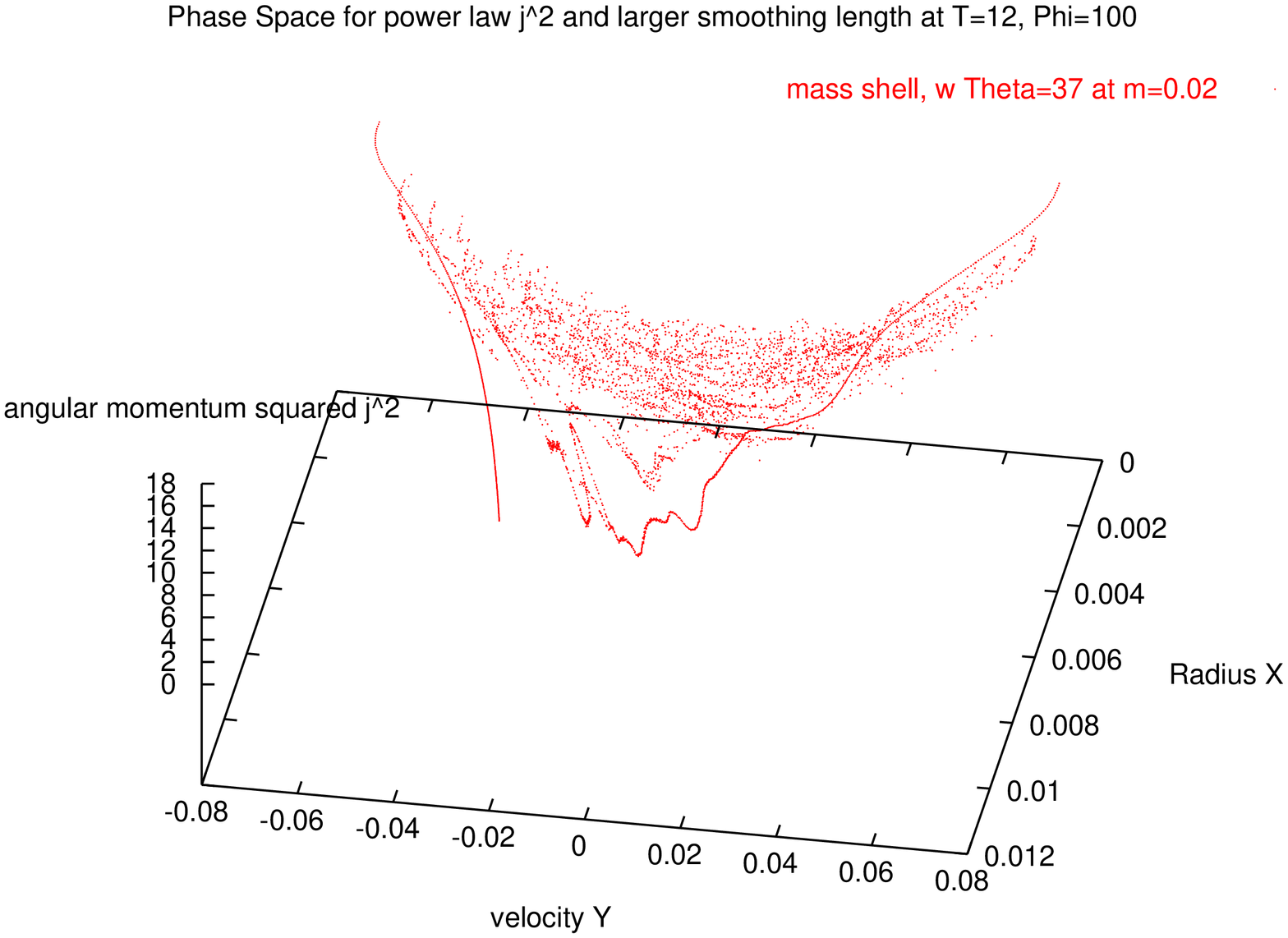}}} \par}

\caption{\label{fig:3DpwrPhSp1.5}Phase space exploration using the `power
law' form of angular momentum near the end of the self-similar quasi-equilibrium
phase (shallow initial density contrasts \protect\( \epsilon =\frac{3}{2}=1.5\protect \)
). 3d views from top left to bottom: front view from underneath; side
view from underneath; front view from above.}
\end{figure*}

Figs. \ref{fig:3DpwrPhSp1.5}, denoted from top left to bottom right
(a), (b) and (c), allow for a finer analysis of the final state: (a) shows a clear relaxation. The core appears
to be rather smoothly populated in this projection. Fig. \ref{fig:3DpwrPhSp1.5}
(b) displays more clearly the outer phase mixing streams. The most striking
fact 
that is made clear in Fig. \ref{fig:3DpwrPhSp1.5} (c) is that the
relaxed  shells lie on a thin surface in phase space: this is a 2-dimensional sheet in the shape of a soaring bird, the neck and beak pointing toward
large radius and angular momentum, while the high velocity wings spread
progressively over higher and higher values of $j^2$. The accumulation of shells at low radius
 forms a flat basin.

\begin{figure*}
{\centering \resizebox*{0.495\textwidth}{!}{\rotatebox{-90}{\includegraphics{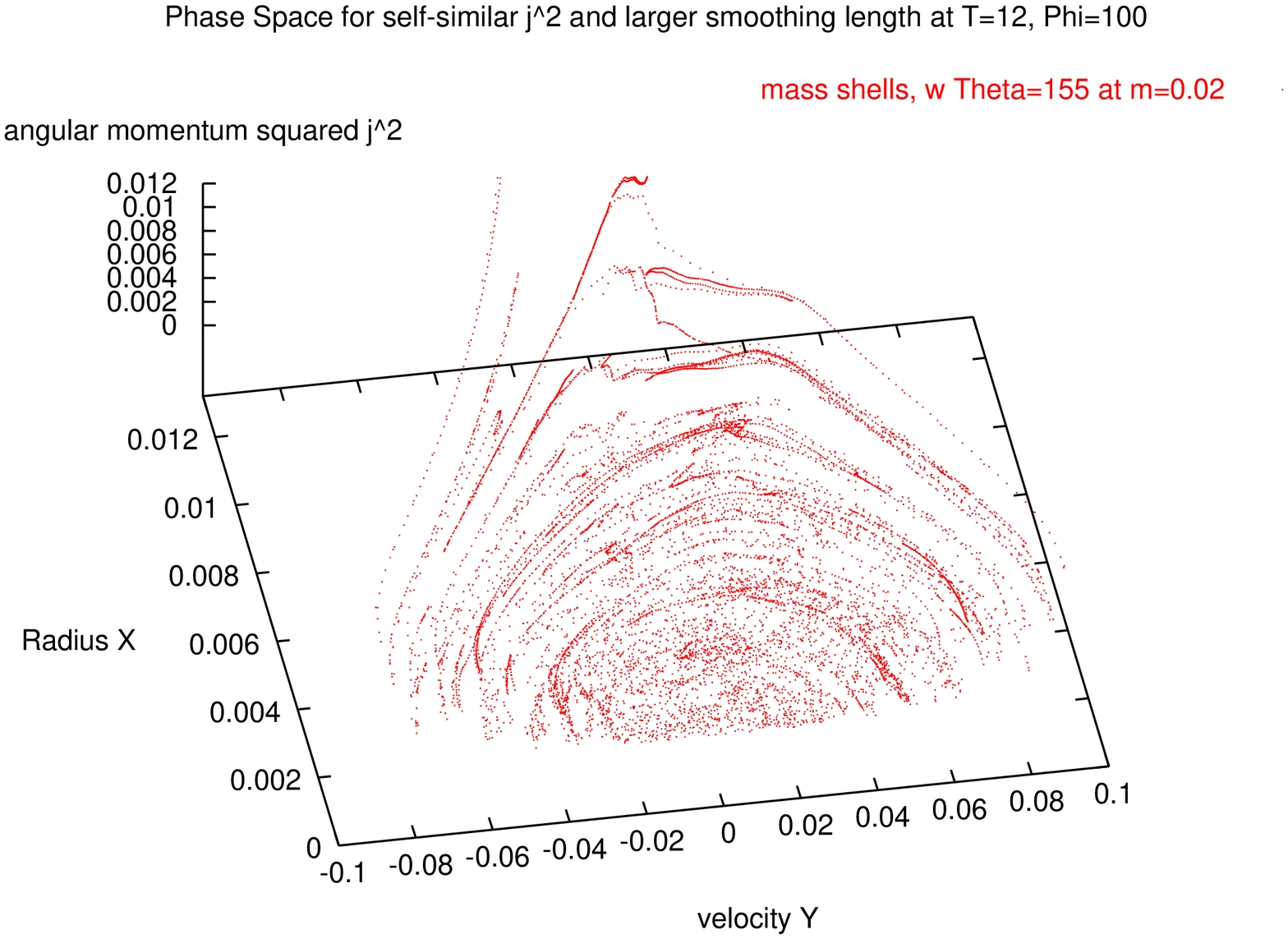}}} \resizebox*{0.495\textwidth}{!}{\rotatebox{-90}{\includegraphics{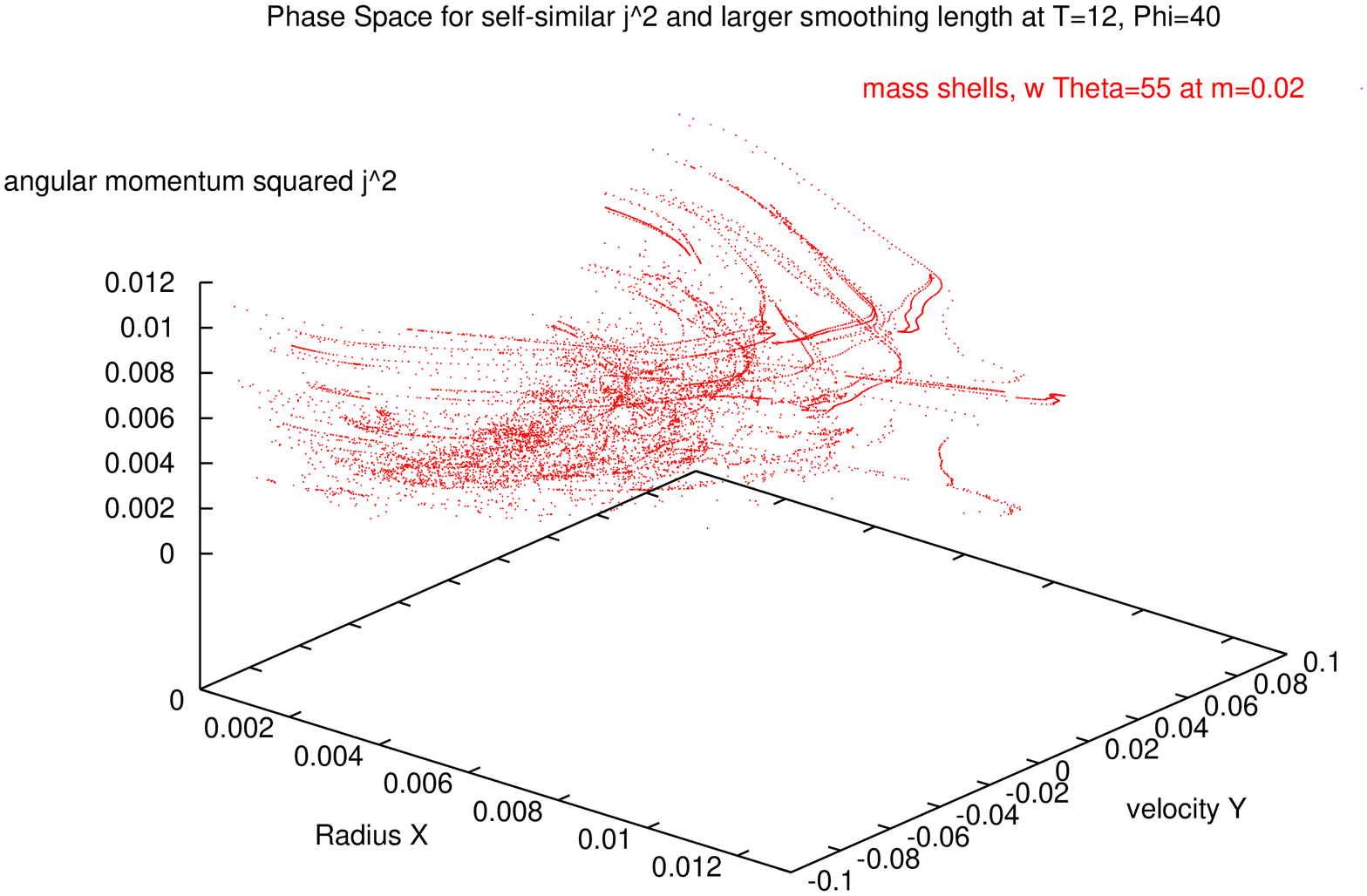}}} \par}

{\centering \resizebox*{0.496\textwidth}{!}{\rotatebox{-90}{\includegraphics{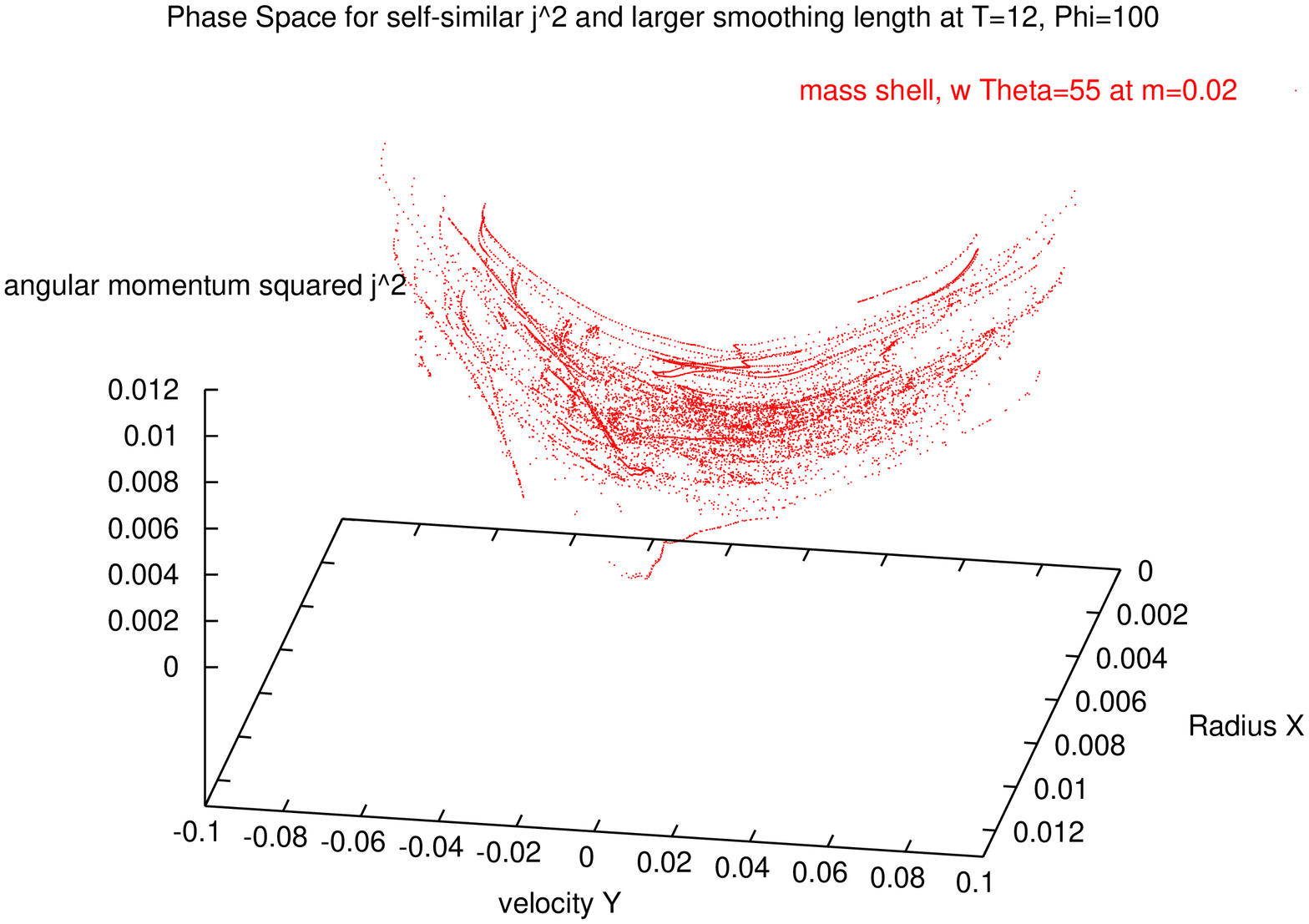}}} \par}

\caption{\label{fig:3Dssj2PhSp1.5}Phase space exploration using the self-similar
form of angular momentum near the end of the self-similar quasi-equilibrium
phase (shallow initial density contrasts \protect\( \epsilon =\frac{3}{2}=1.5\protect \)
). 3d views from top left to bottom: front view from underneath; side
view from underneath; front view from above.}
\end{figure*}

 The phase space from
the SSAM initial angular momentum distribution displays increased 
Liouville stream structure (and is therefore less relaxed) than for
the power law case presented above. Fig. \ref{fig:3Dssj2PhSp1.5}
reveals that the distribution of particles, although it peaks near 
the same surface as that found for the `power law' initial angular momentum
distribution in Fig. \ref{fig:3DpwrPhSp1.5}, is non-zero throughout 
a substantial region of phase space compared to the mean
position. {\it This is more pronounced for the SSAM despite the
  difference in scales used for the angular momentum in the two
  figures}. Relaxation is clearly not as advanced in this case. It
seems that the strict self-similarity before shell-crossing better
segregates the shells in phase space. This recalls the importance of available
phase space volume to relaxation pointed out by Tormen \etal (\cite{Tormen97}),
Teyssier \etal (\cite{Teyssier97}) and Moutarde \etal
(\cite{Moutarde}). The enhanced phase space volume per particle inhibits relaxation. 

\section{Angular momentum and mass correlation?}\label{sec:j2MassCorellationSec}

The existence of the phase space surface found in the previous section
implies correlations in the various projections. Recently one group has reported finding a universal
angular momentum profile in N-body simulations in terms of a halo's
cumulative mass (\cite{Bullock01}) according to \begin{equation}
\label{BullocketalMofJ}
M(<j)=M_{v}m ~j/(j_{0}+j),
\end{equation}
 where \( m  \) and \( j_{0} \) are correlated characteristic
scales and \( M_{v} \) is the halo's virialized mass. However, one 
should bear in mind that the profile given by Eq.(\ref{BullocketalMofJ}) has 
been criticized by other authors (\cite{vandenBosch02,ChenJing}), who claim it 
results from the ommission of particles, carrying what they call negative 
angular momentum, and thus not to represent truly a halo angular momentum 
profile. In fact, their negative angular momentum stands for an angular 
momentum projected on the total vector of the halo which points oppositely to 
the total vector. In our approach one half of the particles carry, according to
this definition, negative angular momentum since our total vector should be 
\( \overrightarrow{0} \) by symmetry. Throughout we have only worked with the 
square of the particle angular momentum.

(\cite{Bullock01}) also find
a power law describing the correlation between angular momentum
and radially cumulative mass (\( j(M)\propto M^{s} \) where \( M=M(<r) \)
and \( s=1.3\pm 0.3 \)). For this reason we consider these
correlations in the two systems of the previous section in order to
determine  whether our choice of initial angular momentum is
compatible with these simulations.

Dimensional analysis  and
the definition of the final density profile as \( \rho \propto r^{-\mu } \)
leads one to approximate the self-similar model's relaxed state  with power
laws for the angular momentum (\( j^{2}\propto r^{4-\mu } \)) and
for the mass profiles (\( M\propto r^{3-\mu } \)). Thus, \( s \)
can be predicted to be \( s=\frac{4-\mu }{2(3-\mu )}, \). In turn the
SSIM gives the index $\mu$ as a function of the initial index $\epsilon$
as:\[
\mu =\left\{ \begin{array}{ll}
2 & \epsilon \leq 2\\
\frac{3\epsilon }{1+\epsilon } & \epsilon >2
\end{array}\right. .\]
Although in the presence of enough angular momentum the radial
SSIM results are expected to be altered, we are not really in that
regime. Thus it is remarkable that
the shallow initial density profile yields
\( s=1. \)  The largest
deviation in the SSIM from that value of \( s \) is given for \( \epsilon =3, \)
since then \( s=\frac{\epsilon +4}{6}=\frac{7}{6}\simeq 1.17, \)
which is also within error from the (\cite{Bullock01}) value.

\begin{figure*}
\vspace{0.1cm}
{\centering \resizebox*{0.9\textwidth}{!}{\includegraphics{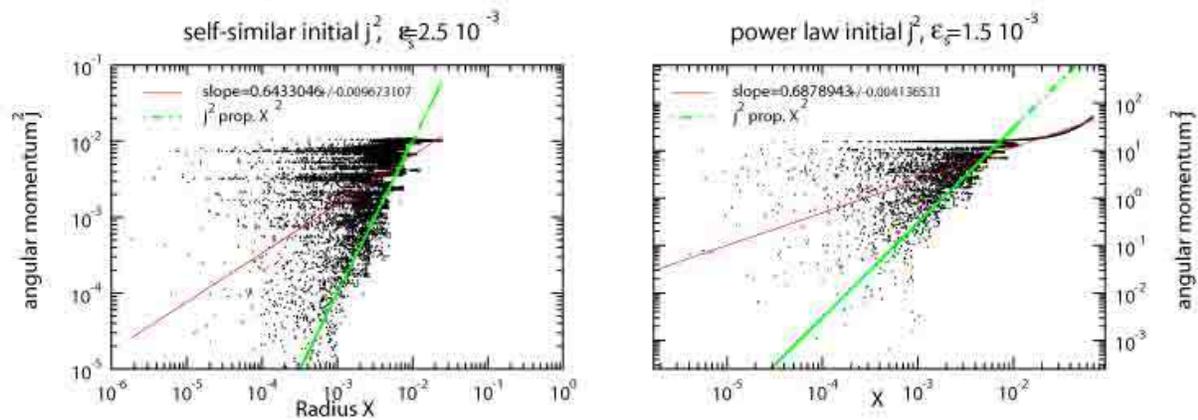}} \par}

\caption{\label{fig:JofX}Two angular momentum profiles and their correlations}
\end{figure*}

Empirically, the previous section's phase space structure warns us
that a correlation \(M\) - \( j^{2} \)  may not be most readily
found from
a simple regression on all shells. Instead we might try to fit just the
correlation from the projected mean surface about which the shells are
distributed.
The projection of phase space into the
\( j^{2}-X \) plane forms a caustic that  appears to be a reasonable
measure to compare with the predicted angular momentum profile. Fig. \ref{fig:JofX} illustrates the angular
momentum profile correlations: the thin continuous lines are mere
power law fits to all of the particles, and their values do
not reflect the self-similar calculations. By contrast, a \( j^{2}\propto X^{2} \)
fit to the phase space caustic verifies the predictions {\it for the
shallow case}. We have not investigated for this note the steep case but have 
no reason to believe the results would be different. Moreover, the density 
profiles show some steepening
on the outer edges of the halo (Fig. \ref{fig:pwrDen}),
which translates into evidence for a flattening of the mass profile,
hence of the mass-angular momentum profile as in Eq.(\ref{BullocketalMofJ}).

These results are remarkable in that they rest on the idea that the
dimensional angular momentum distribution used initially
(\ref{eq:jkep})
also obtains in the ultimate halo. The same thing is true for the
SSAM. Consequently the final correlation $j\propto r$ merely reflects the
universality of the density profile in the intermediate ranges for the
initially `flat' density perturbation.

\section{Summary}
\label{sec:summary}

In this note we have studied the effects of particle angular momentum
on the density profile and phase space structure of the final halo in
the Self-Similar Infall Model (SSIM) for the formation of dark matter
halos. We have used a simple shell code since the angular momentum
averaged over each sphere is zero. We find the NFW profile to be an excellent fit to the density
profile produced when an initial angular momentum distribution
provided by simple dimensional arguments (PLAM) is assigned. This fit holds
until the very central regions where extra flattening is expected on
thermodynamic grounds (this flattening probably proceeds to a Gaussian
with sufficient resolution (\cite{HLD02})).

In addition we have shown that the initial line in phase space
develops into a fairly well defined surface in phase space in the
final halo. The `ridge' or cusp of this surface when projected into
angular momentum- position space is also predicted by the same
dimensional argument. It seems then that some memory of the initial
correlation between position and angular momentum is retained in the
final object. 

An alternate distribution of initial angular momentum (SSAM) that was designed
to maintain strict self-similarity until shell crossing produces a
less relaxed final phase space distribution (at the same dimensionless time) that
is rather less precisely distributed on a 2D surface in the phase
space. It is not able to reproduce the NFW density profile, perhaps
because of the reduced relaxation (as displayed by the higher
dimensionality in the final phase space). 

The question as to whether the successful angular momentum
distribution (PLAM) is actually established by tidal effects remains
unanswered here. It essentially requires the rotational kinetic energy
acquired by each halo particle to be a fixed fraction of the local
halo gravitational potential. That the local gravitational potential
should be an upper limit is clear, but it is not obvious why there
should be a fixed fraction for all shells. One might assign $J^2$ as a
random variable about some mean to see how sensitive the results are
to a fixed value in the manner of (\cite {Sikivie}). 
However a coarse graining over the shells would tend
to produce the same mean independent of $r$ and so we would expect a
similar coarse-grained result to what we have found here.
 In fact
Ryden \& Gunn (\cite{RydenGunn}) do give the expected rms angular
momentum distribution with mass for Gaussian random primordial
perturbations.  Their Fig. 10 suggests a
linear relation in the range of masses that interests us here. This
agrees  approximately with our Eq. (\ref{eq:explj2}) when $\epsilon$ is close
to 2 at moderate radii.

However although the results of the n-body simulations are
usually given as spherical averages, it is not evident that this gives
the same result as the strictly spherically symmetric calculation
reported here. Nevertheless the agreement with the results of Bullock
\etal (\cite{Bullock01}) is encouraging in this sense , even though they may not contain all of the relevant physics(\cite{vandenBosch02,ChenJing}).

\section{Acknowledgments}
RNH acknowledges the support of an operating grant from the canadian
Natural Sciences and Research Council. M LeD wishes to acknowledge the
financial support of Queen's University, Universit\'e Claude Bernard - 
Lyon 1 and Observatoire de Lyon, and to thank St\'ephane Colombi for 
stimulating discussions.

\end{document}

\
\label{lastpage}

\end{document}